\documentclass[twocolumn,aps,superscriptaddress,showpacs,prl]{revtex4}
\usepackage[latin1]{inputenc} 
\usepackage[francais]{babel}
\usepackage{graphicx}
\usepackage{url}
\begin{document}

\NoAutoSpaceBeforeFDP 

\title{Atomic and Electronic Structure of a Rashba $\mbox{\boldmath $p$-$n$}$ Junction at the BiTeI Surface}
\author{C.~Tournier-Colletta}
\email{cedric.tournier@epfl.ch}
\affiliation{Institute of Condensed Matter Physics, Ecole Polytechnique F\'ed\'erale de Lausanne (EPFL), CH-1015 Lausanne, Switzerland}
\author{G.~Aut\`es}
\affiliation{Institute of Theoretical Physics, Ecole Polytechnique F\'ed\'erale de Lausanne (EPFL), CH-1015 Lausanne, Switzerland}
\author{B.~Kierren}
\address{Institut Jean Lamour, UMR 7198,
Nancy-Universit\'e - B.P. 239, F-54506 Vand\oe uvre-l\`es-Nancy,
France}
\author{Ph.~Bugnon}
\affiliation{Institute of Condensed Matter Physics, Ecole Polytechnique F\'ed\'erale de Lausanne (EPFL), CH-1015 Lausanne, Switzerland}
\author{H.~Berger}
\affiliation{Institute of Condensed Matter Physics, Ecole Polytechnique F\'ed\'erale de Lausanne (EPFL), CH-1015 Lausanne, Switzerland}
\author{Y.~Fagot-Revurat}
\address{Institut Jean Lamour, UMR 7198,
Nancy-Universit\'e - B.P. 239, F-54506 Vand\oe uvre-l\`es-Nancy,
France}
\author{O.~V.~Yazyev}
\affiliation{Institute of Theoretical Physics, Ecole Polytechnique F\'ed\'erale de Lausanne (EPFL), CH-1015 Lausanne, Switzerland}
\author{M.~Grioni}
\affiliation{Institute of Condensed Matter Physics, Ecole Polytechnique F\'ed\'erale de Lausanne (EPFL), CH-1015 Lausanne, Switzerland}
\author{D.~Malterre}
\address{Institut Jean Lamour, UMR 7198,
Nancy-Universit\'e - B.P. 239, F-54506 Vand\oe uvre-l\`es-Nancy,
France}
\begin{abstract}
The non-centrosymmetric semiconductor BiTeI exhibits two distinct surface terminations that support spin-split Rashba surface states. Their ambipolarity can be exploited for creating spin-polarized $p$-$n$ junctions at the boundaries between domains with different surface terminations. We use scanning tunneling microscopy/spectroscopy (STM/STS) to locate such junctions and investigate their atomic and electronic properties. The Te- and I-terminated surfaces are identified owing to their distinct chemical reactivity, and an apparent height mismatch of electronic origin. The Rashba surface states are revealed in the STS spectra by the onset of a van Hove singularity at the band edge. Eventually, an electronic depletion is found on interfacial Te atoms, consistent with the formation of a space charge area in typical $p$-$n$ junctions.
\end{abstract}


\maketitle
In inversion-asymmetric systems, the spin-orbit interaction lifts the spin degeneracy. This effect can occur either at surfaces (the Rashba-Bychkov effect~\cite{bych84}) or in the bulk of non-centrosymmetric crystals (the Rashba-Dresselhaus effect~\cite{dress55,rash60}). Such Rashba systems are promising candidates for manipulating electron spin by means of electric field in the context of emerging spintronic devices \cite{datt90}. Significant research efforts are currently directed towards the search of materials exhibiting a ``giant'' Rashba effect that would enable nanometer-scale spintronic devices operating at room temperature. To date, the largest spin splitting, measured by angle-resolved photoemission spectroscopy (ARPES), has been reported in the BiAg$_2$/Ag(111) surface alloy \cite{ast07} and both the surface and bulk states of non-centrosymmetric semiconductor BiTeI \cite{ishi11,saka12,crep12,land12}.

\begin{figure}[b!]
  \includegraphics[width=7.5cm]{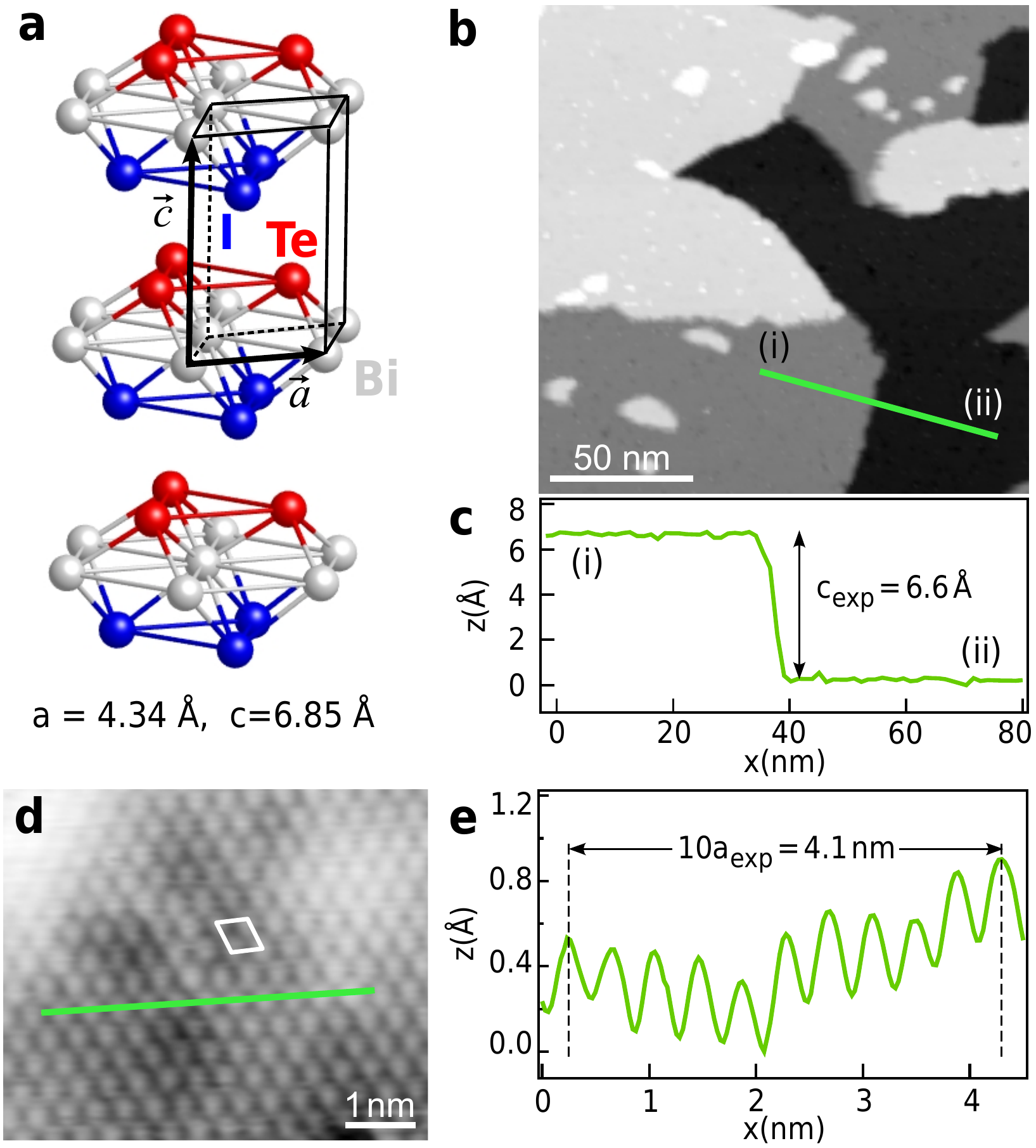}
  \caption{(color online) (a) Layered crystal structure of non-centrosymmetric BiTeI. Trigonal unit cell is depicted by thin lines and unit vectors are shown by arrows. The cleavage plane is between Te (red) and I (blue) atomic planes. (b) STM image covering 170~nm~$\times$~170~nm area shows multiple steps and terraces ($U=-2$~V, $I=50$~pA). (c) Height profile corresponding to the green line in (b). (d) Atomically-resolved STM image ($U=-0.5$~V, $I=20$~pA) with unit cell depicted by the white line. (e) Corrugation profile corresponding to the green line in (d).}
  \label{Fig1}
\end{figure}

In BiTeI, photoemission data show two distinct types of 
surface domains with different surface terminations (Te and I) and opposite surface band bendings that support $p$- or $n$-type surface states \cite{crep12}. We anticipate that Rashba $p$-$n$ junctions can be observed at the boundaries between such domains. This can be used for fulfilling another requirement for fabricating logic devices -- ambipolarity, that is the possibility to control carriers' nature (electrons or holes). Moreover, a number of novel transport phenomena have been predicted for $p$-$n$ junctions
in systems with strong spin-orbit coupling \cite{yama09,taka11,liu12,wang12} that can be further exploited in practical applications. In view of applications, the questions of current dissipation to the bulk or controlled growth of the junction by epitaxy are important, but well beyond the scope of this paper. Here, cleaved BiTeI is rather considered as a toy-system providing a quite unique opportunity to study a $p$-$n$ junction in 2D. 
 
In this work, we employ scanning tunneling microscopy (STM) and spectroscopy (STS) to locate and investigate $p$-$n$ junctions naturally occuring at a cleaved BiTeI surface. The surface domains with Te and I terminations are identified owing to their distinct chemical reactivity, apparent height and electronic structure. The onset of a van Hove singularity at the band edge allows one to identify the Rashba surface states in the STS data. An electronic depletion is observed on interfacial Te atoms, consistent with the picture of a space charge area forming in $p$-$n$ junctions \cite{sze}.

The STM experiments were carried out in a ultra-high vacuum (UHV) setup, using a commercial LT-Omicron scanning tunnelling microscope operated at 5~K. We used  Pt--Ir tips, that are quickly transferred into the vacuum chamber after pinching them off. Spectroscopic $dI/dV$ acquisition is achieved with the lock-in technique, in the open feedback loop mode. Typical modulation bias and frequency were 20~mV and 700~Hz, respectively. The tip's density of states is tailored on Au(111) until a typical Shockley state spectrum is obtained. According to the standard normalization procedure \cite{mart89}, we plot $(dI/dV) / (\overline{I/V})$ to obtain the local density of states (LDOS). $\overline{I/V}$ is obtained by gaussian convolution, that does not affect the peak positions. High-quality single crystals of BiTeI, in the form of platelets, were grown by chemical vapor transport and by the Bridgman technique. Transport measurement show a typical $n$-type degenerate semiconducting behavior. Cleaves were performed at room temperature, using scotch tape, at a pressure of $\approx 10^{-9}$ mbar. The cleaved samples were quickly transferred to the STM head ($P \approx 10^{-10}$ mbar) where they can be measured for several hours. First-principles electronic structure calculations were performed within the density functional theory (DFT) framework employing the generalized gradient approximation (GGA) as implemented in the QUANTUM-ESPRESSO package \cite{gia09}. Spin-orbit effects were accounted for using the fully relativistic norm-conserving pseudopotentials acting on valence electron wavefunctions represented in the two-component spinor form \cite{dal05}. The surface band structures were obtained using a slab model with Te- and I-terminated surfaces. 

The BiTeI crystal has a trigonal layered structure ($a=4.34$~\AA, $c=6.85$~\AA\ \cite{shev94}) consisting of trilayers stacked along the $c$ axis [Fig.~\ref{Fig1}(a)]. Within a trilayer, Bi and Te are covalently bonded to form a positively charged (BiTe)$^{+}$ bilayer. The interaction between the latter and the I$^{-}$ atomic planes is of ionic nature. The weakest bonding between the Te and I atomic planes belonging to adjacent trilayers defines the natural cleavage plane. As a result of its non-centrosymmetric structure, two sides of the cleave have different surface terminations. However, due to the presence of stacking faults in the bulk of the crystal both Te- and I-terminated domains can be observed on a single surface \cite{crep12}. 

The typical surface morphology after cleaving is shown in Figure~\ref{Fig1}(b). We observe terraces with a typical width of several tenths of nanometers. In Figure~\ref{Fig1}(c), the profile taken between points (i) and (ii) shows a step height of 6.6~\AA\ which is in agreement with the bulk lattice constant. The BiTeI structure is finally confirmed by examining the atomically-resolved STM image in Figure~\ref{Fig1}(d). We indeed observe an in-plane hexagonal atomic pattern. From corrugation profile [Fig.~\ref{Fig1}(e)], we obtain a lattice constant of $4.1$~\AA\ that again compares well with the diffraction data. 

\begin{figure}[b]
  \includegraphics[width=8.7cm]{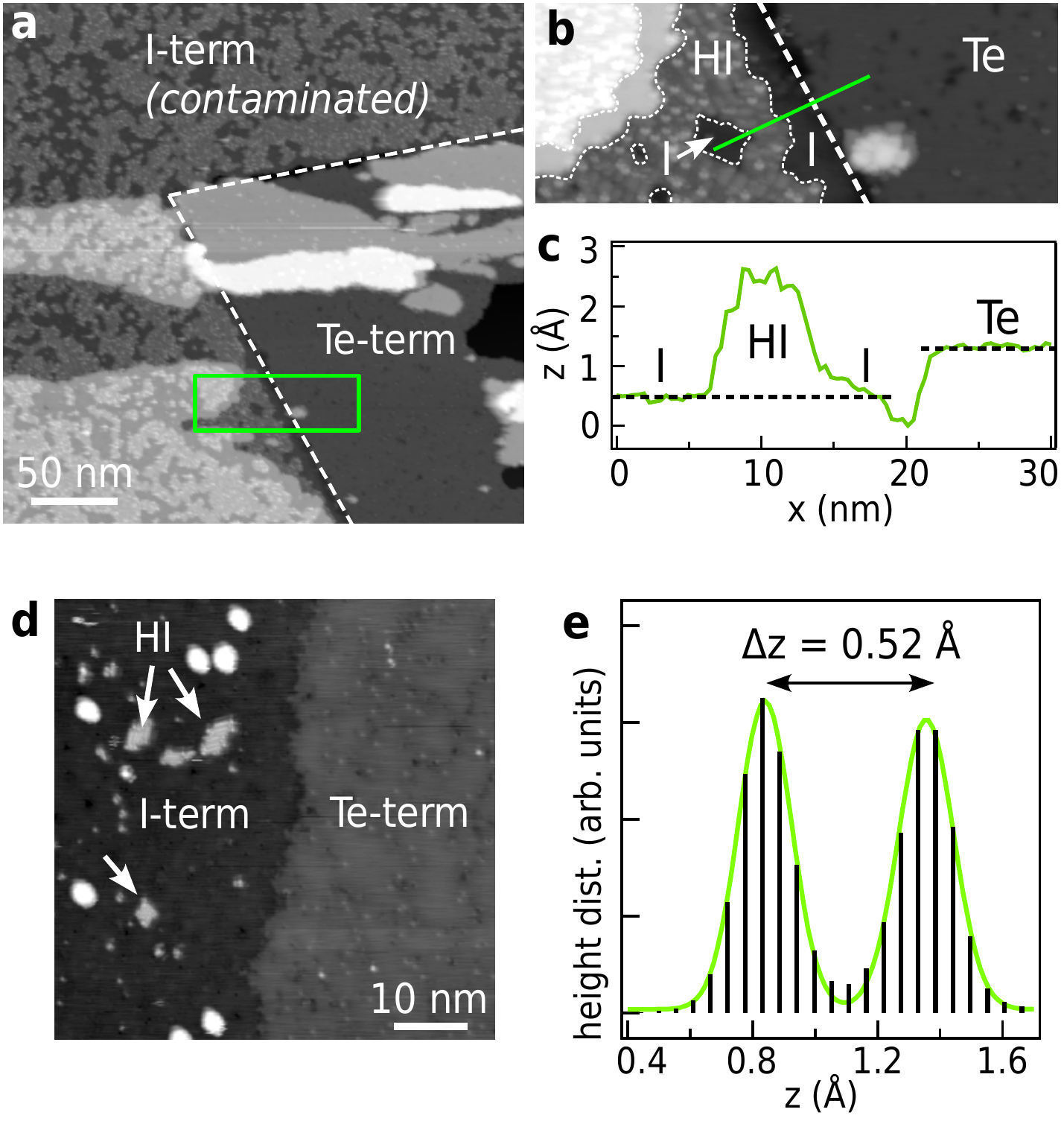}
  \caption{(color online) (a) STM image measured after H$_2$ dosing (1 L), showing the higher reactivity of the I-terminated surface. (b) Close-up corresponding to green rectangle in (a). (c) Height profile across the interface, passing through an amorphous HI island. (d) STM image and (e) corresponding $z$-distribution, confirming the height mismatch at a clean surface termination domain boundary.}
  \label{Fig2}
\end{figure}

The surface terminations remain to be identified. The first feature that allows one to distinguish between the two terminations is related to the surface reactivity. Halogen atoms are indeed known to be very reactive, and especially prone to be reduced by hydrogen. To verify that hypothesis, we dosed a freshly cleaved sample by approximately 1~Langmuir of H$_2$. Figure~\ref{Fig2}(a) shows a large-scale image ($U=-2$~V, $I=50$~pA) measured after dosing. We identify the Te and I terminations, judging by the cleanliness of the former and the high degree of contamination of the latter. The contaminated surface is almost completely covered by islands that are attributed to the formation of HI. A close-up image of the interface is shown in Figure~\ref{Fig2}(b). In Figure~\ref{Fig2}(c), we show a height profile measured across the interface [green line in Fig.~\ref{Fig2}(b)], passing through an HI island with a typical height of $\approx 2$~\AA. A height mismatch of 0.75~\AA\ is observed between the clean I- and Te-terminated areas, the latter appearing higher. This result is corroborated for a freshly cleaved sample [Fig.~\ref{Fig2}(d); $U=-3$~V, $I=0.15$~nA] where the I termination, identifiable due to the presence of a few HI islands, appears lower. The corresponding height distribution [Fig.~\ref{Fig2}(e)] shows a mismatch of 0.52~\AA, a smaller value than found previously. Owing to the distinct tunneling parameters, this finding suggests a spectroscopic origin for the observed height mismatch, as demonstrated below in the text. 

\begin{figure}[b]
  \includegraphics[width=8.6cm]{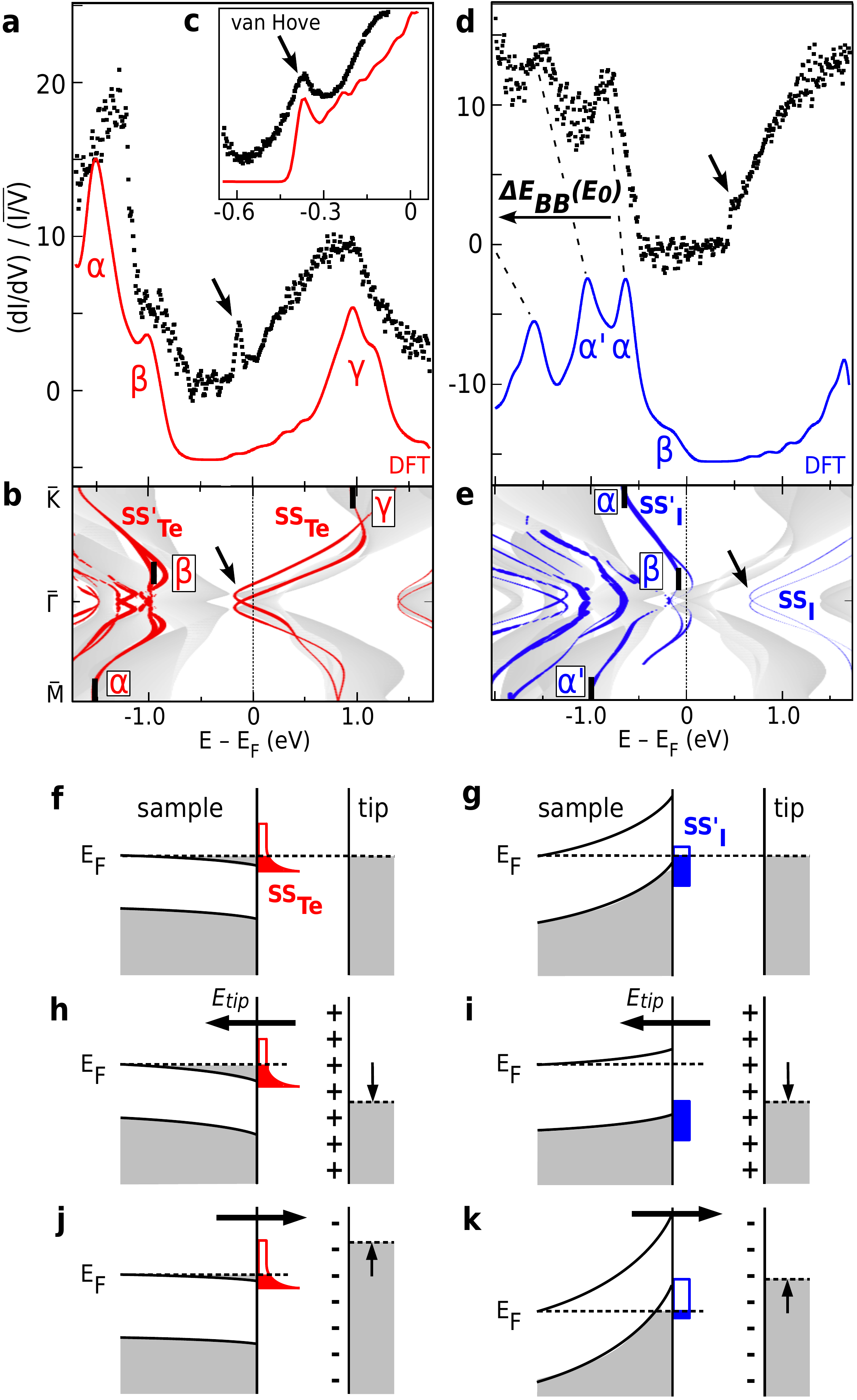}
  \caption{(color online) (a) Normalized $dI/dV$ curve measured for the Te-terminated surface of BiTeI. The solid line which corresponds to the calculated LDOS projected on the surface atom was aligned to match the experimentally observed features. (b) Calculated band structure 
of a slab model superimposed against the bulk band structure projected onto the surface Brillouin zone (shaded area). The line thickness reflects the projected weight of wavefunction on the surface atoms.
(c) Close-up emphasizing the van Hove singularity associated with the Rashba surface state on the Te termination. (d) Normalized $dI/dV$ curve, calculated LDOS and (e) the slab band structure for the I-terminated surface.  Schematic description of the tip-induced band bending: (f,g) junction at equilibrium ($V_{\rm tip}=0$); probing (h,i) occupied states ($V_{\rm tip} >0$) and (j,k) unoccupied states ($V_{\rm tip} <0$). See text for details.}
  \label{Fig3}
\end{figure}

In addition to their different chemical reactivity, the local electronic structure measured by STS provides an independent way of identifying the two terminations. Figure~\ref{Fig3}(a) shows a typical $dI/dV$ spectrum measured on the Te termination of a freshly cleaved sample (setpoint $U=-2$~V, $I=0.4$~nA; ten spectra averaged). The curve is compared to the LDOS on the surface Te atoms for this termination calculated using a 13-trilayer slab model (red line). The latter exhibits three main structures labeled $\alpha$, $\beta$ and $\gamma$, that nicely reproduce the overall shape and peak positions of the experimentally measured spectrum. By examining the corresponding band structure, the peaks are assigned to the extrema of the Rashba surface states [labeled as SS$^{'}_{\rm Te}$ ($\alpha$, $\beta$) and SS$_{\rm Te}$ ($\gamma$) and indicated by vertical lines in Fig.~\ref{Fig3}(b)]. SS$_{\rm Te}$ yields an electron pocket with a band edge located $-0.15$~eV below the Fermi level, in relative agreement with the ARPES experiments  \cite{ishi11,crep12,land12}. 

At this precise energy, the experimental data show an additional, sharp peak indicated by the arrow [Fig.~\ref{Fig3}(a)]. This feature indicates the onset of a 1D-like van Hove singularity at the band edge, a hallmark of the Rashba effect in a two-dimensional electron gas \cite{ast07bis}. Experimentally, such a singularity appears as a finite-width peak due to the finite energy resolution. Reproducing this feature in calculations requires a dense $k$-point sampling around $\bar{\Gamma}$ which necessitates the use of a smaller slab model. Figure~\ref{Fig3}(c) compares the LDOS of a 7-trilayer slab and a high-statistics experiment ($U=-1$~V, $I=0.1$~nA) in a narrow energy range. The calculated DOS is convoluted to account for the experimental resolution ($\delta E \approx 1.7~eV_{\rm mod} = 35$~meV \cite{cram05}). A clear peak now emerges in the calculated LDOS, in agreement with the experiment. However, due to the finite quasiparticle lifetime, the experimental linewidth ($\approx 150$~meV, comparable to ARPES data) is significantly broader than the calculated value. Scattering due to electron-electron inteactions is predicted to yield a contribution of a few meV only \cite{erem12}, therefore the linewidth must be dominated by scattering by phonons and impurities.

Figures~\ref{Fig3}(d,e) show the measured and calculated data for the I-terminated surface. The calculated surface bands are shifted by $\approx 0.9$~eV to lower binding energies compared to the Te-terminated surface due to the sign change of the surface potential \cite{erem12bis}. The Fermi surface is now constituted by a hole pocket associated with the SS$^{'}_{\rm I}$ feature, as confirmed by ARPES \cite{crep12}. As for the case of Te-terminated surface, we can link the main features $\alpha$, $\alpha'$ and $\beta$ in the calculated LDOS [blue line in Fig.~\ref{Fig3}(d)] to the band structure. Peaks $\alpha$ and $\alpha'$ can be recognized in the experimental $dI/dV$ curve. Peak $\beta$ is not clearly identified in the data. According to the calculated LDOS, it should be much weaker than peak $\alpha$. Experimentally, it might be not resolved from peak $\alpha$, but responsible for the asymmetry of the latter. Finally, a close inspection of the $dI/dV$ data at 0.5~eV shows a peak-dip-hump shape [arrow in Fig.~\ref{Fig3}(d)]. Similar to the case of Te-terminated surface, this feature is attributed to the onset of a van Hove singularity at the bottom of the SS$_{\rm I}$ Rashba bands. However, the peak is less pronounced because of a smaller Rashba splitting.

Despite the overall good agreement between experiment and theory, a discrepancy remains as to the gap amplitude. The data on I-terminated BiTeI show a gap of $\approx 1$~eV centred at $E_{\rm F}$ [Fig.~\ref{Fig3}(d)], that is incompatible with (i) the picture of a hole pocket and (ii) the gap amplitude of $\approx 0.6$~eV measured on the Te-terminated surface [Fig.~\ref{Fig3}(a)]. As commonly observed in tunneling experiments on semiconductors, a tip-induced band bending \cite{mcel93} alters the absolute energy scale and complicates the interpretation of the spectra, especially at low temperatures. In particular, the experimental gaps are usually overestimated \cite{feen06,mysl06}. Figures~\ref{Fig3}(f,g) show the tip-sample energy diagram of the junction at equilibrium ($V_{\rm tip}=0$). Occupied bulk states, SS$_{\rm Te}$ and SS$^{'}_{\rm I}$ are depicted in gray, red and blue, respectively. Opposite band bendings characterize the two terminations, with the onset of either an accumulation [Te, Fig.~\ref{Fig3}(f)] or a depletion layer [I, Fig.~\ref{Fig3}(g)] \cite{crep12}. 

When tunneling from sample to tip [$V_{\rm tip} >0$, Fig.~\ref{Fig3}(h,i)], the tip's positive charge produces an electric field that penetrates into the semiconductor and produces a downward band bending \footnote{At low temperature, the concomitant accumulation of holes at the surface would produce a field oriented in the same direction.}. At the Te-terminated surface [Fig.~\ref{Fig3}(h)], the effect is small because a slight bending brings enough bulk carriers at the surface to screen the electric field. Therefore, the measured and calculated features are comparable. However at the I-terminated surface [Fig.~\ref{Fig3}(i)], a large band bending can occur until the conduction band is populated. As a result, the experimental features appear at energies $E_{exp}$ that exceed the calculated binding energies $E_0$. Assuming a linear relation $E_{0} = E_{\rm exp} (1-x)$ for peaks $\alpha$ and $\alpha'$, we obtain $x \approx 0.6$, a value similar to that found at the Si(111)-$7\times7$ surface \cite{mysl06}. In turn, we estimate that peak $\beta$ should shift by 0.2~eV, which explains the absence of states at $E_{\rm F}$ and the seemingly larger gap in the measurements performed on I-terminated surface of BiTeI. When tunneling from tip to sample [$V_{\rm tip} <0$, Fig.~\ref{Fig3}(j,k)], an upward band bending is produced. For the same reason as before, the effect is very weak in the Te case [Fig.~\ref{Fig3}(j)]. This is also true in the I case [Fig.~\ref{Fig3}(k)], because an accumulation of holes occurs at the top of the valence band at low voltages. These holes strongly screen the tip's negative charge at higher voltages \cite{feen06}.

\begin{figure}[t]
  \includegraphics[width=8.7cm]{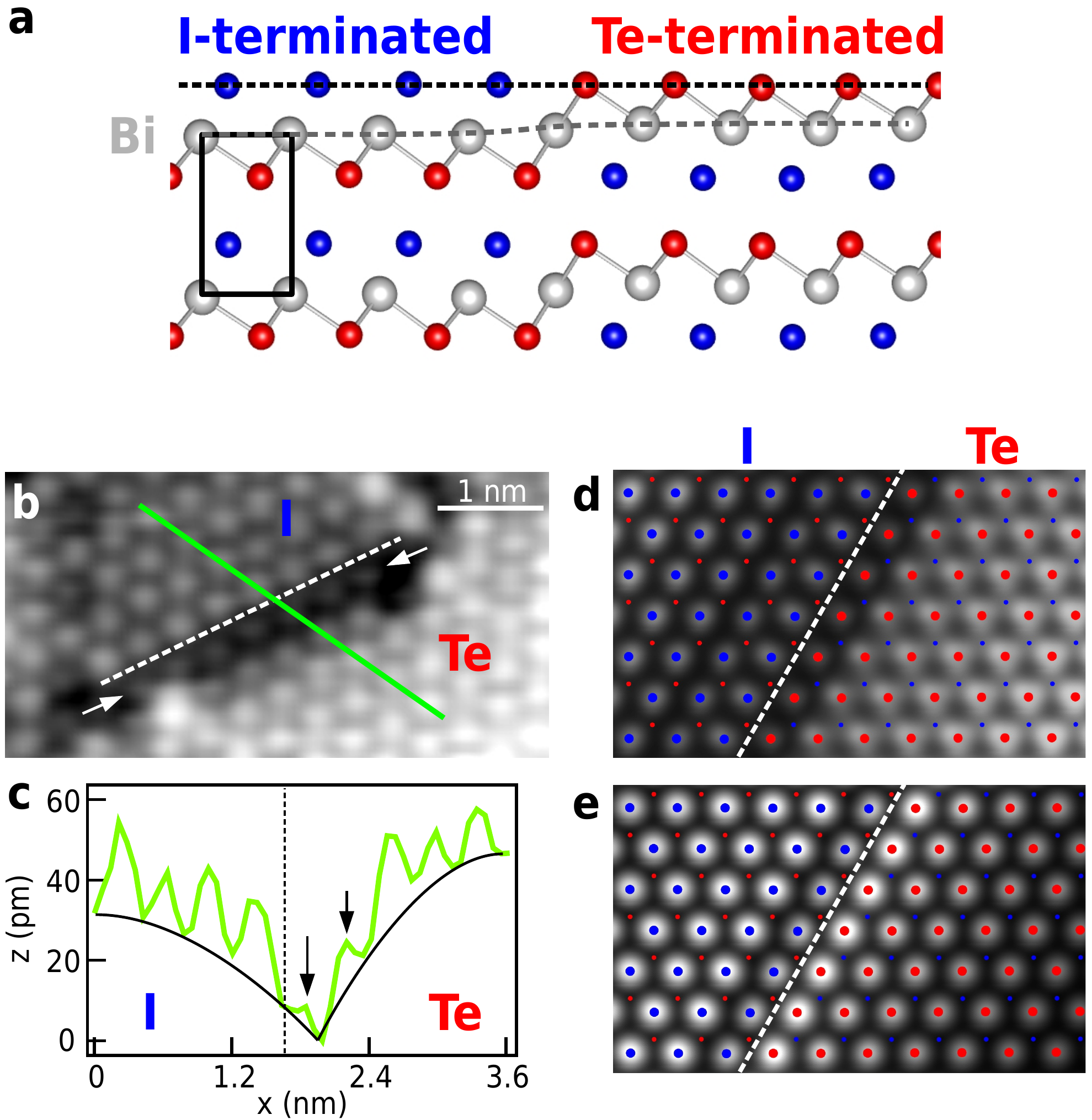}
  \caption{(color online) (a) Relaxed atomic structure of the surface termination domain boundary viewed along the interface direction. The bulk unit cell is indicated by rectangle. (b) STM image of the interface ($U=-1.8$~V, $I=0.1$~nA) showing atomic resolution for both surface terminations (separated by dashed line). (c) Corrugation profile corresponding to green line in (b) evidence the charge depletion on interfacial Te atoms. The plain line is a guide to the eye showing a typical width of 2$-$3 nm for the space charge area. (d,e) Calculated STM images probing filled ($U=-1.8$~V) and empty ($U=1.8$~V) states, respectively.}
  \label{Fig4}
\end{figure}

Finally, we examine the atomic-scale details of the boundary formed by the Te- and I-terminated domains. Figure~\ref{Fig4}(a) shows the structure calculated using a bulk supercell model consisting of alternating domains of 10$a \approx$~40~\AA\ width. We started from a bulk configuration where Bi planes had same height on Te- and I-terminated domains. Upon relaxation, the Bi planes tend to bend resulting in practically the same height of Te and I atomic planes throughout the bulk. In the calculations performed on a slab model of the interface, the atomic coordinates remain essentially the same. Therefore, the height mismatch pointed out in Fig.~\ref{Fig2}(c) cannot be related to any step formation at the domain boundary. The STM image calculated for the relaxed structure [Fig.~\ref{Fig4}(d)] reproduces the apparent contrast observed experimentally in the occupied states [Fig.~\ref{Fig4}(b); $U=-1.8$~V, $I=0.1$~nA], hence demonstrating the electronic origin of this feature. The observed contrast is at odds with the naive expectation that the iodine atoms should appear brighter due to their higher electronegativity. Actually, the contrast is attributed to the surface states and it appears that the wavefunction maximum is below (above) the surface plane at the I (Te) termination \cite{erem12bis}.

It is interesting to note that both experiment [Fig.~\ref{Fig4}(b),(c)] and calculations [Fig.~\ref{Fig4}(d)] show also a clear depletion on the Te surface atoms at the interface. Since no structural distorsion is seen in the calculation, we again conclude an electronic origin of this phenomenon. Calculations actually show that the overall contrast is inverted upon probing the empty states [Fig.~\ref{Fig4}(e)], with the I termination appearing higher. Right at the boundary, the Te atoms now appear as protusions while their I counterparts are slightly depleted. This feature is consistent with the picture of a space charge region typical for $p$-$n$ junctions. In our case, electrons are transfered from the Te to the I termination that act as $n$- and $p$-type surface semiconductors, respectively. Interestingly, the space charge area shows a very narrow width of 2$-$3 nm that compares well with the calculated band bending length scale \cite{ishi11,erem12bis}. This property could be exploited to fabricate devices of a few nanometers only, to be compared to the $\approx$30~nm length scale of the present-day CMOS technology.


In conclusion, we combined STM/STS measurements and DFT calculations to investigate the atomic and electronic properties of the domain boundary formed by Te- and I-terminated BiTeI. We showed that electronic effects strongly affect the topographic measurements, in particular by the formation of a space charge area typical of $p$-$n$ junctions. The relevant surface states are characterized by a strong spin-orbit coupling, as demonstrated by the van Hove singularity in the STS data. This system thus provides a realization of a 2D Rashba $p$-$n$ junction, where surface states play the role of doped carriers in analogy with standard bulk semiconductor devices. More experimental work remains to be done, in particular to carry out transport experiments with a four-point probe. Eventually, the distinct surface state binding energy in related compounds BiTeBr and BiTeCl \cite{erem12,saka13,land13,chen13} offers an interesting way to manipulate the Rashba carrier concentration.

\begin{acknowledgements}
The authors acknowledge helpful discussions with J. C. Johannsen, A. Crepaldi and L. Moreschini. G.A. and O.V.Y were supported by the Swiss NSF grant No.~PP00P2\_133552 and the ERC starting grant ``TopoMat'' (No.~306504). First-principles computations have been performed at the Swiss National Supercomputing Centre (CSCS) under project s443. 
\end{acknowledgements}


\end{document}